\documentclass[twocolumn,showpacs,preprintnumbers,amsmath,amssymb]{revtex4}
\usepackage{tabularx,graphicx}\begin{document}
\newcommand{\beq}{\begin{equation}}
\newcommand{\eeq}{\end{equation}}
\newcommand{\beqn}{\begin{eqnarray}}
\newcommand{\eeqn}{\end{eqnarray}}
\newcommand{\bmath}{\begin{subequations}}
\newcommand{\emath}{\end{subequations}}
\title{Dynamic Hubbard model: kinetic energy driven charge expulsion, charge inhomogeneity, hole superconductivity, and Meissner effect}
\author{J. E. Hirsch }
\address{Department of Physics, University of California, San Diego\\
La Jolla, CA 92093-0319}
 
\date{\today} 
\begin{abstract} 
Conventional Hubbard models do not take into account the   fact that the wavefunction of an electron in an atomic orbital expands when a second electron occupies the orbital.
Dynamic Hubbard models have been proposed to describe this physics.   These models reflect the fact that electronic materials are generically $not$ electron-hole symmetric, and they
give rise to superconductivity driven by lowering of kinetic energy when the electronic 
energy band is almost full, with higher transition temperatures
resulting when the ions are negatively charged. 
We show  that the   charge distribution in dynamic Hubbard models can be highly inhomogeneous in the presence of disorder, and that a finite system will expel 
{\it negative charge} from the interior to the surface, and that these tendencies are largest 
in the parameter regime where the models give rise to highest superconducting transition temperatures.
High $T_c$ cuprate materials exhibit charge inhomogeneity and they exhibit tunneling asymmetry, a larger tendency to emit electrons rather than holes in
NIS tunnel junctions.
We propose that these properties, as well as their high $T_c$'s, are evidence that they are governed by  the physics described by dynamic Hubbard models. 
 Below the superconducting transition temperature
the models considered here describe a negatively charged superfluid and positively charged quasiparticles, unlike the situation in conventional
BCS superconductors where quasiparticles are charge neutral on average. We examine the temperature dependence of
the superfluid and quasiparticle charges and conclude that spontaneous electric fields should be observable in the interior and in the vicinity of superconducting
materials described by these models 
at sufficiently low temperatures. We furthermore suggest that the dynamics of the negatively charged superfluid and positively charged
quasiparticles in dynamic Hubbard models can provide an explanation for the Meissner effect observed  in high $T_c$ and low $T_c$  
superconducting materials.
\end{abstract}
\maketitle 

\section{Introduction}
The wavefunction of electrons in an  atom is self-consistently determined by all the electrons in the atom\cite{orbx}. The conventional single band Hubbard  Hamiltonian\cite{hub,hub2}
\beq
H=-\sum_{i,j,\sigma}[t_{ij}c_{i\sigma}^\dagger c_{j\sigma}+h.c.]+U\sum_i n_{i\uparrow}n_{i\downarrow}
\eeq
does not take this well-known fact into account: the atom with two electrons is assumed to change its energy due to electron-electron repulsion, but the
electronic  wavefunction is assumed
to be simply the product of single-electron wavefunctions. This is incorrect because the spacing of electronic energy levels in an atom is always smaller than the 
electron-electron repulsion\cite{inapp}. To correct this deficiency we have proposed a variety of new Hamiltonians\cite{holeelec}, generically called `dynamic Hubbard models', that take into account the fact
that {\it orbital expansion} takes place when a non-degenerate atomic orbital is 
doubly occupied. These Hamiltonians involve either an auxiliary spin\cite{hole1,tang} or boson degree of freedom\cite{dynhub}, or a second electronic orbital\cite{hole2}.

A simple way to incorporate this physics is by the site Hamiltonian\cite{pincus,dynhub}
\beq
H_i=\frac{p_i^2}{2m}+\frac{1}{2} K q_i^2+(U+\alpha q_i)n_{i\uparrow}n_{i\downarrow}
\eeq
where $\alpha$ is a coupling constant (assumed positive) and $q_i$ a local boson degree of freedom  describing the orbital relaxation,
with equilibrium position at $q_i=0$ if zero or one electrons are present.  The Coulomb repulsion between electrons is $U$ when $q_i=0$.
However, upon double occupancy of the orbital at site $i$, $q_i$ will take the value $q_i=-\alpha/K$
and the on-site repulsion will be reduced from $U$ to $U_{eff}=U-\alpha^2/(2K)$
to give rise to   minimum energy, as can be seen from completing the square:
\beq
H_i=\frac{p_i^2}{2m}+\frac{1}{2} K( q_i+\frac{\alpha}{K}n_{i\uparrow}n_{i\downarrow})^2+(U-\frac{\alpha^2}{2K})n_{i\uparrow}n_{i\downarrow} .
\eeq
This is a way to describe the orbital relaxation\cite{relax} that takes place when the orbital becomes doubly occupied.
The conventional Hubbard model corresponds to the limit of an infinitely stiff orbital ($K\rightarrow \infty$) where the orbital does
not relax and the on-site $U$ is not reduced.

The importance of this physics increases when the ionic charge (positive) is small\cite{dynhub}, since in that
case the orbital expansion is larger (for example, the orbital expansion is larger for $H^-$ than for $He$), which corresponds to a smaller
stiffness parameter $K$ in Eq. (3).  The importance of this physics for a lattice system of such atoms also increases as the filling of the electronic energy band increases and
there is an increasing number of atoms with doubly occupied orbitals. For these two reasons, the importance of this physics
increases the more {\it negative charge} the system has. 
Thus it is reasonable to expect that a lattice system described by this model will have a strong
tendency to {\it expel negative charge}\cite{chargeexp0,chargeexp}.  

Figure 1 depicts the essential physics of dynamic Hubbard models as opposed to conventional Hubbard models: the doubly occupied orbital is larger
than the singly occupied orbital. Note also that 
when an atomic orbital expands, the electronic kinetic energy is lowered, since in an orbital of radial extent  $r$ the electron kinetic energy is of order $\hbar^2/(2m_e r^2)$, 
with $m_e$ the electron mass. Thus, one can say that in the atom as described by a dynamic Hubbard model there is {\it negative charge expulsion driven by 
kinetic energy lowering}. This is not the case for an atom described by the conventional Hubbard model.

  \begin{figure}
\resizebox{8.5cm}{!}{\includegraphics[width=7cm]{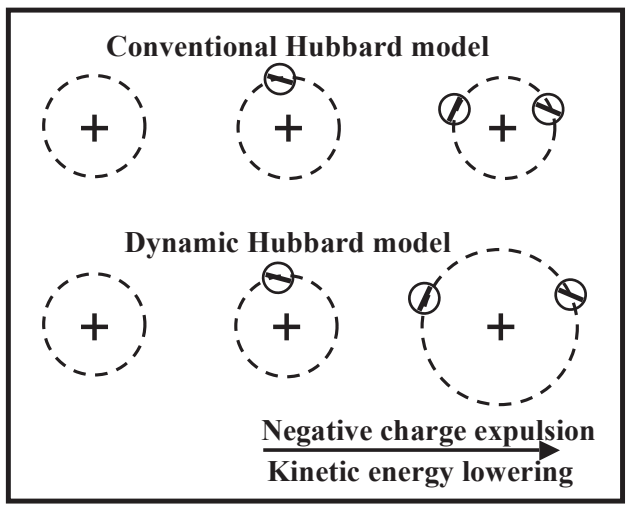}}
  \caption{In the conventional Hubbard model the atomic orbital is not modified by electronic occupancy. In the dynamic Hubbard model
  and in real atoms, addition of
  the second electron causes orbital expansion due to the electron-electron interaction. Negative charge is expelled outward and the kinetic energy
  of the electrons is lowered relative to that   with a non-expanded orbital.  }
\end{figure}

Just like for the atom, we find for the system as a whole described by a dynamic
Hubbard model that there is negative charge
expulsion and it is associated with lowering of kinetic energy.
Figure 2 shows results of exact diagonalization for a finite lattice that indicate
that  the electron
concentration is considerably larger near the surface than in the interior. We
discuss the origin of this effect and the details of the calculation leading to the results shown in Fig. 2 in the following sections.

 \begin{figure}
\resizebox{5.5cm}{!}{\includegraphics[width=7cm]{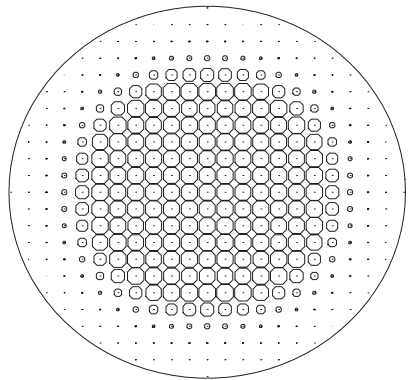}}
  \caption{Cylindrical superconductor described by the dynamic Hubbard model. 
The circles at each lattice site have radius proportional to the hole occupation at the site. Note that the hole occupation is larger in the interior than near
the surface, implying that the negative charge concentration is higher near the surface than in the interior. }
\end{figure} 

 \begin{figure}
\resizebox{6.0cm}{!}{\includegraphics[width=7cm]{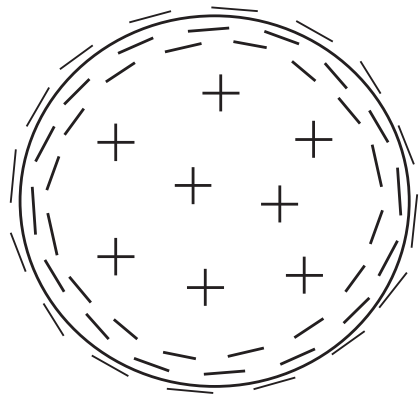}}
  \caption{Expected charge distribution in the ground state of superconductors according to the theory of hole superconductivity\cite{chargeexp}.
  The superconducting condensation energy and associated charge expulsion originate in lowering of kinetic energy.}
\end{figure} 

The theory of hole superconductivity\cite{website,sns} starts from a dynamic Hubbard model and predicts within BCS theory\cite{deltat} a superconducting state with some essential differences
from the conventional superconducting state. The superconducting condensation energy originates in lowering of $kinetic$ rather 
than potential energy\cite{apparent,99,prb2000}, and the gap function is energy-dependent with a slope of universal sign\cite{deltat}. 
Also, superconductors are predicted to have a non-homogeneous charge distribution in the ground state\cite{chargeexp0}, as shown
schematically in Fig. 3: excess positive charge in the interior and excess negative charge near the surface, resembling a ``giant atom''\cite{giantatom}. 
This charge distribution is predicted from modified London-like electrodynamic equations\cite{chargeexp}, as well as from the 
energy dependence of the gap function\cite{chargeexp0}. 
The charge distribution in Fig. 3 is $qualitatively$ similar to that in Figs. 2 and 1. 

The finite gap slope of the superconducting gap function
predicted by dynamic Hubbard models\cite{deltat}  leads to the  prediction of asymmetric tunneling characteristics in NIS tunnel junctions\cite{tunasym}, with larger current for a negatively
biased superconductor, reflecting the tendency of the superconductor to expel electrons. Such asymmetric behavior of NIS tunnel junctions is commonly found in high
temperature superconductors.

As a consequence of the charge expulsion physics the superconducting state in systems described by dynamic Hubbard models has quasiparticles that are $positively$ charged on average\cite{thermo}, and   the superfluid has excess $negative$ charge, in contrast
to conventional BCS-London superconductors where quasiparticles are charge neutral on average. We will examine the consequences of this physics
for superconductors described by these models at temperatures well below the superconducting transition temperature.

Dynamic Hubbard models are by nature electron-hole $asymmetric$ and so are superconductors, as evidenced by the fact that a rotating superconductor develops a magnetic field that is always {\it parallel}, never antiparallel, to the direction of the mechanical angular
momentum of the body\cite{ehasym}. This suggests that dynamic Hubbard models are  more appropriate to describe superconductors than the conventional Hubbard
model that is electron-hole symmetric. Furthermore, in dynamic Hubbard models kinetic energy lowering plays a key role, in contrast to conventional Hubbard models. 
It was pointed out  early on by Fritz London that\cite{londonkinetic} ``the superfluid state of helium as well as the superconducting state of electrons are fluid'' (rather than solid), and that
this may arise because ``it will be more favorable to give preference to minimizing the kinetic energy''.
In his books\cite{londonbooks}, London emphasized this physics for superfluid $^4He$ but not 
for superconductors. We have recently pointed out the
close relationship between the physics of superconductors as described by the
theory of hole superconductivity and superfluid $^4He$\cite{super2}.

A non-uniform charge distribution in a solid gives rise to
electrostatic fields and an associated potential energy cost. It will be favored if this cost is compensated by a kinetic energy gain, i.e. lowering of kinetic energy.
Thus it is natural to expect  that dynamic Hubbard models are prone to develop charge inhomogeneity, 
and in extreme cases phase separation\cite{prb}, where the kinetic energy lowering overcompensates for the potential energy cost. High $T_c$ cuprates  exhibit charge inhomogeneity\cite{bianconi, stripes,patches,patches2}, suggesting that dynamic Hubbard models may be
useful to describe them.

In superconductors described by the conventional BCS-London theory, no negative charge expulsion occurs, the kinetic energy is raised rather than
lowered in the transition to
superconductivity, quasiparticles are charge neutral on average, and the Meissner effect is argued to be completely
understood  within the framework of the conventional theory\cite{m1,m2,m3,m4,m5,m6,m7,m8,m9,m10,m11}. 
However, we have argued elsewhere that the conventional theory does not provide a $dynamical$ understanding of the Meissner effect\cite{lorentz,validity}. Instead, the negative charge expulsion physics driven by kinetic energy lowering 
of dynamic Hubbard models discussed here offers a natural explanation for
the Meissner effect\cite{emf,meissner}: just as in classical plasmas obeying Alfven's theorem\cite{plasma}, the magnetic field lines
move with the expelled negative charge. The physics of dynamic Hubbard models is proposed to apply to all superconducting materials,
in contrast to the conventional theory that is proposed to apply only to ``conventional'' superconductors\cite{cohen,norman}. Given that $all$ superconductors
exhibit a Meissner effect, it is useful to remember Isaac Newton's rule of natural philosophy\cite{newton}:
{\it  to the same natural effects we must, as far as possible, assign the same causes.''}

\section{dynamic Hubbard models}

We can describe the physics of interest by a multi-orbital tight binding model (at least two orbitals per site)\cite{hole2,multi2}, or with a background 
spin\cite{hole1,color} or harmonic oscillator\cite{phonon1,phonon2} degree of freedom that is coupled to the electronic double occupancy, as in Eq. (2). Assuming the latter, the   site Hamiltonian is given by Eq. (2),
and the Hamiltonian can be written as
\beqn
H&=&-\sum_{i,j,\sigma}[t_{ij}c_{i\sigma}^\dagger c_{j\sigma}+h.c.]+ \sum_i \hbar \omega_0 a_i^\dagger a_i\nonumber \\
&+&\sum_i [U+g\hbar \omega_0)(a_i^\dagger + a_i)]n_{i\uparrow}n_{i\downarrow}
\eeqn
with frequency $\omega_0=\sqrt{K/m}$ and $g=\alpha/(2K\hbar\omega_0)^{1/2}$ the dimensionless coupling constant. Estimates for the values of these parameters were discussed in
ref. \cite{dynhub}.  
Using a generalized Lang-Firsov transformation\cite{mahan,dynhub,phonon2,hawai} the electron creation operator $c_{i\sigma}^\dagger$ is written in terms of new quasiparticle operators
$\tilde{c}_{i\sigma}^\dagger$ as
\beqn
c_{i\sigma}^\dagger&=&e^{g(a_i^\dagger-a_i)\tilde{n}_{i,-\sigma}}\tilde{c}_{i\sigma}^\dagger=[1+(e^{-g^2/2}-1)\tilde{n}_{i,-\sigma}]\tilde{c}_{i\sigma}^\dagger \nonumber \\
&+&\tilde{n}_{i,-\sigma} \times (incoherent \;  part)
\eeqn
where the incoherent part describes the processes where the boson goes into an excited state when the electron is created at the site. 
For large $\omega_0$ those terms become small and   we will ignore them in what follows, which amounts to keeping only ground state to ground state
transitions of the boson field.
The electron creation operator is then given by
\bmath
\beq
c_{i\sigma}^\dagger=[1+(S-1)\tilde{n}_{i,-\sigma} ]\tilde{c}_{i\sigma}^\dagger
\eeq
\beq
S=e^{-g^2/2}
\eeq
and the quasiparticle weight for electronic band filling $n$ ($n$ electrons per site) is
\beq
z(n)=(1+(S-1)\frac{n}{2})^2
\eeq
\emath
so that it decreases monotonically from $1$ when the band is almost empty to $S^2<1$ when the band is almost full. 
The single particle Green's function and associated spectral function is renormalized by the multiplicative factors on the quasiparticle operators given
 in Eq. (6a))\cite{phonon2,undr}, which on the average amounts to multiplication of the spectral function
by the quasiparticle weight Eq. (6c). 
This will cause a reduction in the photoemission spectral weight at low energies from what would naively follow from the low energy effective Hamiltonian,
an effect extensively discussed in Ref. \cite{undr}. A corresponding reduction occurs in the
two-particle Green's function and associated low frequency optical properties\cite{undr,dynhub12}.

The low energy effective Hamiltonian is then
\bmath
\beq
H=-\sum_{ij\sigma}  t_{ij}^\sigma [\tilde{c}_{i\sigma}^\dagger \tilde{c}_{j\sigma}+h.c.]+U_{eff}\sum_i \tilde{n}_{i\uparrow}\tilde{n}_{i\downarrow}
\eeq
\beq
t_{ij}^\sigma=[1+(S-1)\tilde{n}_{i,-\sigma} ][1+(S-1)\tilde{n}_{j,-\sigma} ] t_{ij}
\eeq
\emath
and $U_{eff}=U-\hbar \omega_0 g^2$. Thus, the hopping amplitude for an electron between sites $i$ and $j$ is given by $t_{ij}$, $St_{ij}$ and $S^2t_{ij}$ depending on whether there
are $0$, $1$ or $2$ other electrons of opposite spin at the two sites involved in the hopping process.

The physics of these models is determined by the magnitude of the  parameter $S$, which can be understood as the overlap matrix element between the
expanded and unexpanded orbital in Fig. 1. It  depends crucially on the net ionic charge $Z$, defined as the
ionic charge when the orbital in question is unoccupied\cite{dynhub}. In Fig. 1, $Z=1$ if the 
states depicted correspond to the hydrogen ions $H^+$, $H$ and $H^-$ and  $Z=2$ if 
they correspond to $He^{++}$, $He^+$ and $He$.  In a lattice of $O^=$ anions, as in the $Cu-O$ planes of high $T_c$ cuprates, 
the states under consideration are $O$, $O^-$ and $O^=$ and $Z=0$,
and in the $B^-$ planes of $MgB_2$, $Z=1$. The effects under consideration here become larger when $S$ is small, hence when $Z$ is small.
An approximate calculation of $S$ as a function of $Z$ is given in \cite{dynhub}.

We now perform a particle-hole transformation since we will be interested in the regime of low hole concentration. The hole creation operator is  given by, instead of Eq. (6a)
\bmath
\beq
c_{i\sigma}^\dagger=[S+(1-S)\tilde{n}_{i,-\sigma} ]\tilde{c}_{i\sigma}^\dagger
\eeq
where $\tilde{n}_{i,\sigma}$ is now the hole site occupation, and the hole quasiparticle weight increases with hole occupation $n$ as
\beq
z_h(n)=S^2(1+(\frac{1}{S}-1)\frac{n}{2})^2
\eeq
\emath

For simplicity of notation we   denote the hole creation operators again by $c_{i\sigma}^\dagger$,
the hole site occupation by $n_{i\sigma}$  and the 
effective on-site repulsion between holes of opposite spin $U_{eff}$ (the same as between electrons)  by $U$ to simplify the notation. The Hamiltonian for holes is then
\bmath
\beq
H=-\sum_{ij\sigma}  t_{ij}^\sigma [ {c}_{i\sigma}^\dagger  {c}_{j\sigma}+h.c.]+U \sum_i  {n}_{i\uparrow} {n}_{i\downarrow}
\eeq
\beq
t_{ij}^\sigma=t_{ij}^h[1+(\frac{1}{S}-1)) n_{i,-\sigma} ][1+(\frac{1}{S}-1){n}_{j,-\sigma} ] 
\eeq
\emath
with $t_{ij}^h=S^2t_{ij}$ the hopping amplitude for a single hole when there are no other holes in the two sites involved in the hopping process. The hole hopping amplitude
and the effective bandwidth  increase
as the hole occupation increases, and so does the quasiparticle (quasihole) weight Eq. (8b).

Finally, we will assume there is only nearest neighbor hopping $t_{ij}=t$  for simplicity and write the nearest neighbor hopping amplitude 
resulting from Eq. (9b) as
\bmath
\beq
t_{ij}^\sigma=t_h+\Delta t(n_{i,-\sigma}+n_{j,-\sigma})+\Delta t_2 n_{i,-\sigma}n_{j,-\sigma}
\eeq
with
\beq
t_h=tS^2
\eeq
\beq
\Delta t=tS(1-S)
\eeq
\beq
\Delta t_2=t(1-S)^2=(\Delta t)^2/t_h   .
\eeq
\emath
The non-linear term with coefficient $\Delta t_2$ is expected to have a small effect when the band is close to full (with electrons)  and is often neglected.
Without that term, the model is also called the generalized Hubbard model or Hubbard model with correlated hopping\cite{kiv,camp}.
The effective hopping amplitude for average site occupation $n$ is, from Eq. (10a)
\beq
t(n)=t_h+n\Delta t+\frac{n^2}{4}\Delta t_2
\eeq
so that a key consequence of integrating out the higher energy degrees of freedom is to renormalize the hopping amplitude and hence the 
bandwidth and the effective mass (inverse of hopping amplitude).

Generalized dynamic Hubbard models which include also coupling of the boson degree of freedom to the single site occupation have qualitatively similar physics,
since the low energy effective Hamiltonian is also given by Eq. (7). They 
are discussed in Ref. \cite{undr}.

\section{hole pairing and superconductivity in dynamic Hubbard models}
As we\cite{deltat} and others\cite{oth,oth2,oth3,oth4} have discussed, the correlated hopping $\Delta t$ gives a strong tendency  to  pairing and superconductivity when a band
is almost full. The hopping amplitude for a single hole is $t_h$, and it increases to $t_h+\Delta t$ when the hole hops to or from a site occupied by another hole
(of opposite spin), thus giving an incentive for holes to pair to lower their kinetic energy.
The superconductivity described by this model has a number of interesting  features\cite{deltat} that we have proposed are relevant to the
description of high $T_c$ cuprates, namely strong dependence of $T_c$ on hole concentration\cite{deltat}, energy dependent gap function and resulting
tunneling asymmetry of universal sign\cite{tunasym},  superconductivity driven by kinetic energy lowering and associated low energy optical sum rule violation\cite{apparent},
change in optical spectral weight at frequencies much higher than the superconducting energy gap upon onset of superconductivity\cite{color}, strong positive pressure
dependence of $T_c$\cite{deltat}, increased quasiparticle weight upon entering the superconducting state\cite{undr}, etc. Many of these predictions are supported by 
observations on high $T_c$ cuprates made both before and after the predictions were made. 

 The more fundamental dynamic Hubbard model from which the $\Delta t$ interaction derives has also been studied using Eliashberg theory\cite{dynfrank} and
exact numerical methods\cite{dynhub12} and shows an even stronger tendency to pairing and superconductivity.

\section{negative charge expulsion in the normal state of dynamic Hubbard models}

We consider the Hamiltonian for holes Eq. (9), with the hopping amplitudes given by Eq. (10). We assume a cylindrical geometry of radius R and infinite length in the z direction. We decouple the interaction terms within a simple mean field approximation 
assuming $<n_{i\sigma}>=n_i/2$ with $n_i$ the hole occupation at site $i$, and obtain the mean field Hamiltonian
\bmath
\beq
H_{mf}=H_{mf,kin}+H_{mf,pot}
\eeq
\beq
H_{mf,kin}=-\sum_{<ij>,\sigma}[t_h+\Delta t n_i+\Delta t_2 \frac{n_i^2}{4}] [c_{i\sigma}^\dagger c_{j\sigma}+h.c.]
 \eeq

 \beq
H_{mf,pot}=\frac{U}{4}  \sum_i n_i^2
\eeq
\emath
Assuming a band filling of $n$ holes per site, we diagonalize the Hamiltonian Eq. (12) with initial values $n_i=n$ and fill the lowest energy levels until   the occupation $n$ is achieved.
From the Slater determinant of that state we obtain new values of $n_i$ for each site, and iterate this procedure until self-consistently is achieved.  We can  extend this
procedure to finite temperatures, iterating to self-consistency for a given chemical potential $\mu$. 
We consider then the resulting occupation of the sites as function of the distance $r$ to the center of the cylinder. Sometimes there are non-equivalent sites at the same distance from the
axis (e.g. (5,0) and (3,4)) that yield somewhat different occupation, for those cases we show the average and standard deviation as error bars in the graphs.

Figure 4 shows a typical example of the behavior found. Here we assumed $\Delta t_2=0$, corresponding to the  simpler   Hubbard model with
correlated hopping and no six-fermion operator  term. Even for $\Delta t=0$ the hole occupation is somewhat larger in the interior than near the surface.
When the interaction $\Delta t$ is turned on, the hole occupation increases in the interior and decreases near the surface. This indicates that the system expels
electrons from the interior to the surface. Clearly, this occurs because the sites near the surface
have lower coordination than those in the interior and thus benefit less from the lowering of kinetic energy associated with higher hole concentration
(described by Eq. (12b) than the sites in the bulk. The effect becomes more pronounced when $\Delta t$ is increased, as one would expect.

Figure 5 shows the hole site occupations as circles of diameter proportional to it, for the cases $\Delta t=0$ and $\Delta t=0.25$ of Fig. 4. 
Note that the interior hole occupation is   larger for $\Delta t=0.25$ than it is for $\Delta t=0$, while near the surface the hole
occupation  is larger for $\Delta t=0$.  Again this shows that the system with $\Delta t=0.25$ is expelling electrons from the interior
to the surface, thus depleting the hole occupation near the surface.

  \begin{figure}
\resizebox{8.5cm}{!}{\includegraphics[width=7cm]{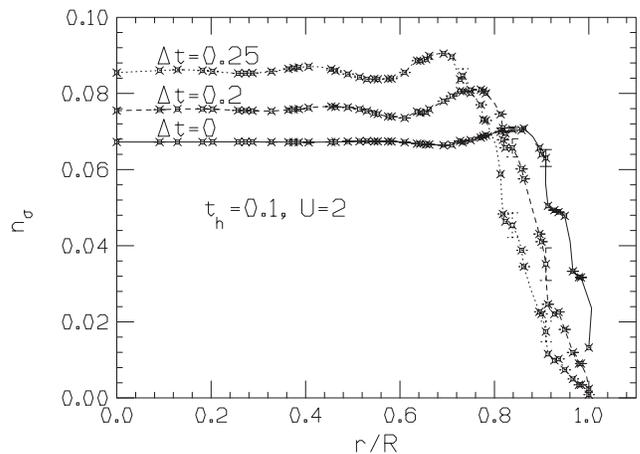}}
  \caption{Hole site occupation per spin for a cylinder of radius $R=11$ as function of $r/R$, with $r$ the distance to the center,
  for a cubic lattice of side length $1$ in the normal state. There are $377$ sites in a cross-sectional area ($\pi R^2=380.1$). The
  average occupation (both spins) is $n=0.126$ holes per site and the temperature is $k_BT=0.02$.}
\end{figure} 

  \begin{figure}
\resizebox{9.0cm}{!}{\includegraphics[width=7cm]{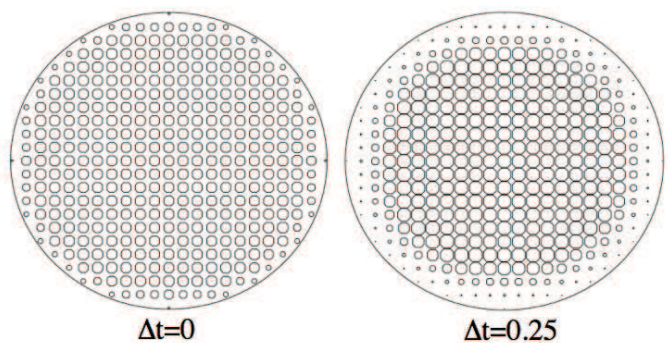}}
  \caption{The diameters of the circles are proportional to the hole occupation of the site. Note that for finite $\Delta t$ the hole occupation increases in the
  interior and is depleted near the surface. The parameters correspond to the cases shown in Fig. 4.}
\end{figure} 

These results are obtained by iteration. Fig. 6 shows the behavior of the energies as a function of iteration number for the cases $\Delta t=0$ and $\Delta t=0.25$ of Fig. 4.
The initial values correspond to a uniform hole distribution with each site having the average occupation. The evolution is non-monotonic because in the
intermediate steps the overall hole concentration increases, nevertheless it can be
seen that for the case $\Delta t=0.25$ the final kinetic energy when self-consistency is achieved is lower, and the final potential energy is higher, associated with the larger
hole concentration in the interior and the lower hole concentration near the surface shown in Fig. 4. This is of course what is expected. For the case $\Delta t=0$ instead
there is essentially no difference
in the energies between the initial uniform state and the final self-consistent state.

  \begin{figure}
\resizebox{8.5cm}{!}{\includegraphics[width=7cm]{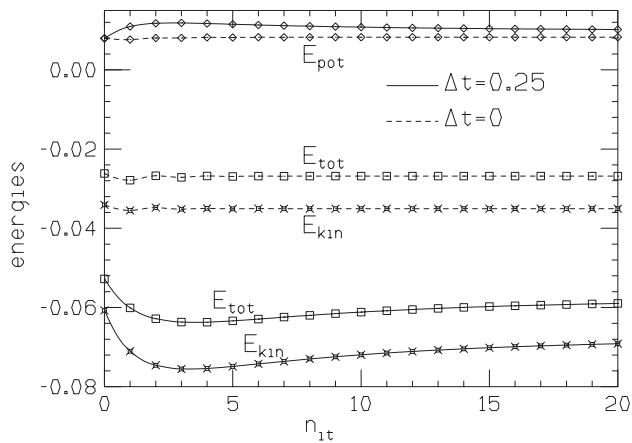}}
  \caption{Kinetic, potential and total energy per site for $\Delta t=0.25$ as function of number of iterations starting with a uniform hole distribution.}
\end{figure}

As the correlated hopping amplitude $\Delta t$ increases, and even more so in the presence of $\Delta t_2$, the system appears to develop a tendency to phase
separation, where holes condense in the interior and the outer region of the cylinder becomes essentially empty of holes. This happens very rapidly as function
of the parameters for the finite system under consideration. Examples are shown in Fig. 7.  
An analytic derivation of the condition on the parameters in the Hamiltonian and
band filling where this occurs is given in ref. \cite{prb}.

  \begin{figure}
\resizebox{8.5cm}{!}{\includegraphics[width=7cm]{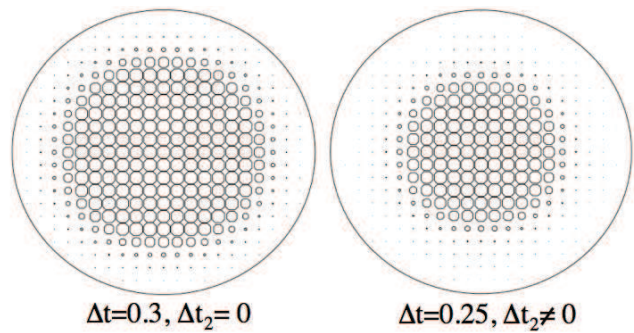}}
  \caption{As the correlated hopping terms increase, the system develops a tendency to phase separation, where essentially all the holes condense to the
  interior. Parameters are the same as in Fig. 4 except as indicated. The maximum hole occupation per spin is  0.128 and 0.214 for the left and right panel, the average hole occupation
  per spin  is 0.063.}
\end{figure}

In summary, we have seen that the dynamic Hubbard model promotes expulsion of negative charge from the interior to the surface of the system in the normal state when the band is almost full,
and that the charge expulsion physics is associated with kinetic energy lowering, just as in the single atom, Fig. 1. When the concentration of holes increases, in other words when the band
has fewer electrons, the negative charge expulsion tendency rapidly decreases\cite{prb}.
The charge expulsion tendency is largest when the parameter $\Delta t$ is largest, which in turn corresponds to smaller $S$, the overlap of the 
atomic orbitals when one and two electrons are at the orbital. As discussed earlier, $S$ is smaller when the ionic charge $Z$ is smaller, corresponding to a 
more negatively charged ion. The fact that the effective Hamiltonian derived from this physics expels more negative charge the more negatively
charged the ion is and the more electrons there are
in the band   makes of course a lot of sense and can be regarded as an internal consistency check on the validity of the model. The largest charge expulsion tendency, occuring  when
$\Delta t$ is large and when the band is close to full, corresponds to the regime giving
rise to highest superconducting transition temperature\cite{deltat}.

For a normal metal, the charge expulsion physics will be compensated to a large extent by  
longer range Coulomb repulsion, since no electric field can exist in the interior of a normal metal.
Nevertheless  as we argue in the next sections some residual effects of charge expulsion can be seen even in the normal state.
For the superconducting state, we have proposed new electrodynamic equations that give rise to ``charge rigidity''\cite{rigidity} and the inability of the superfluid to screen
interior electric fields so that the charge expulsion physics can manifest itself\cite{chargeexp}.

\section{charge inhomogeneity in dynamic hubbard models}

Small local potential variations have a large effect in  dynamic Hubbard models. 
We have shown before that in the superconducting state of the model  there is
great sensitivity to local potential variations due to the slope of the gap function\cite{local}. Here we find that the model is also sensitive to local potential variations in the 
normal state.
Because kinetic energy dominates the physics of the dynamic Hubbard model, the system
will develop charge inhomogeneity at a cost in potential energy if it can thereby lower its kinetic energy more, unlike models where the
dominant physics is potential (correlation) energy like the conventional Hubbard model.

We assume there are impurities in the system that change the local potential at some sites, and compare the effect of such perturbations for the dynamic and
conventional Hubbard models. As an example we take parameters $t_h=0.1$, $U=2$ and consider site impurity potentials of magnitude
$\pm 0.2$ at several sites as indicated in the caption of Fig. 8. For the dynamic Hubbard model we take $\Delta t=0.2$, $\Delta t_2=0.4$, corresponding to
$S=0.333$. 

Figure 8 shows the effect of these impurities on the charge occupation for the conventional and dynamic models. In the conventional Hubbard model the occupation
 changes at the site of the impurity potential and only very slightly at neighboring sites. In the dynamic Hubbard model
the local occupation change at the impurity
site itself is much larger than in the conventional model, and in addition,   the occupations
change at many other sites in the vicinity of the impurities, as seen in the lower panel of Figure 8. Figure 9 shows the real space distribution
of these changes.

 \begin{figure}
\resizebox{8.5cm}{!}{\includegraphics[width=7cm]{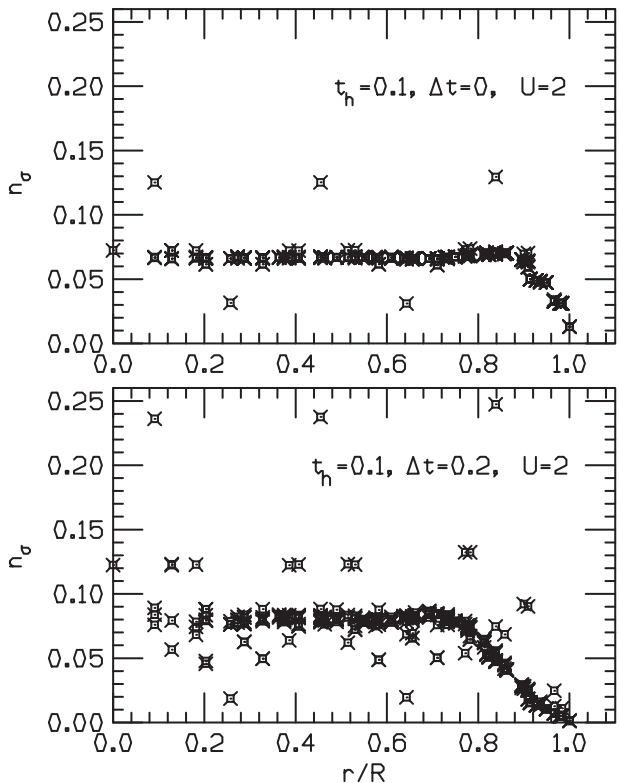}}
  \caption{Hole site occupation per spin in the normal state in a system of radius $R=11$ with 5 impurities at positions
  (-1,0), (2,2), (3,-4), (-5, -5), (-6, 7) with potential strength   -0.2, +0.2, -0.2, +0.2, -0.2 respectively. Note the much larger variation in densities
  generated in the dynamic Hubbard model (lower panel, $\Delta t_2\neq 0$) than in the conventional Hubbard model (upper panel).
  Average hole occupation per site is $n=0.126$.}
\end{figure}

  \begin{figure}
\resizebox{8.5cm}{!}{\includegraphics[width=7cm]{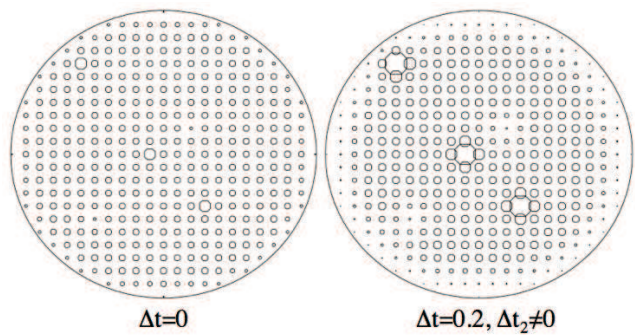}}
  \caption{The site occupations for the case of Fig. 8, with the diameters of the circles proportional to the hole occupation of the sites.
  Note the 5 impurity sites at positions listed in the caption of Fig. 8 (three with negative potential, hence higher hole concentration) and two with positive potential, hence lower hole concentration.
  Note that for $\Delta t=0$ only the occupation at the impurity site changes appreciably, while for $\Delta t \neq 0$ an impurity potential of the
  same strength causes a much larger change of the occupation at the impurity site and   occupation   change also at the nearest
  and next nearest neighbor sites.}
\end{figure}

The reason for this large sensitivity to local perturbations can be understood from the form of the hopping amplitude Eq. (9b). A change in the local
occupation will also change the hopping amplitude of a hole between that site and neighboring sites, which in turn will change the occupation of 
neighboring sites, and so on.  In that way a local perturbation in the dynamic Hubbard model
gets amplified and expanded to its neighboring region, and it is easy to understand how a non-perfect crystal will easily develop patches of
charge inhomogeneity in the presence of small perturbations. These inhomogeneities cost potential (electrostatic) energy,
but are advantageous in kinetic energy. The conventional Hubbard model does not exhibit this physics.
There is extensive experimental evidence for
charge inhomogeneities in high $T_c$ cuprates\cite{bianconi,stripes,patches,patches2}.

 \section{shape effects}

It is interesting to consider the effect of the shape of the sample on the charge expulsion profile in the dynamic Hubbard model. Consider an ellipsoidal shape as shown
in Figure 10. The sites near the surface at the regions of higher curvature, i.e. top and bottom, have somewhat smaller hole concentration than at the regions of lower curvature at the lateral surfaces.
This is easy to understand: the sites near the surface in the regions of high surface curvature have slightly lower coordination on average  than those in the regions of low curvature, hence
the holes do not benefit so much from kinetic energy lowering and prefer to stay away from those regions. Thinking in terms of electrons instead of holes, it means
the body expels more electrons to the top and bottom than to the sides. This should give rise to a higher electric potential near the sides than at the top and bottom,
and a quadrupolar electric field with field lines starting at the lateral sides and ending at the top and bottom. This is precisely the type of electric field
found in the superconducting state by solving the alternative London equations proposed to describe the electrodynamics of superconductors within this
theory\cite{chargeexp,ellipsoid}.

\begin{figure}
\resizebox{7.5cm}{!}{\includegraphics[width=7cm]{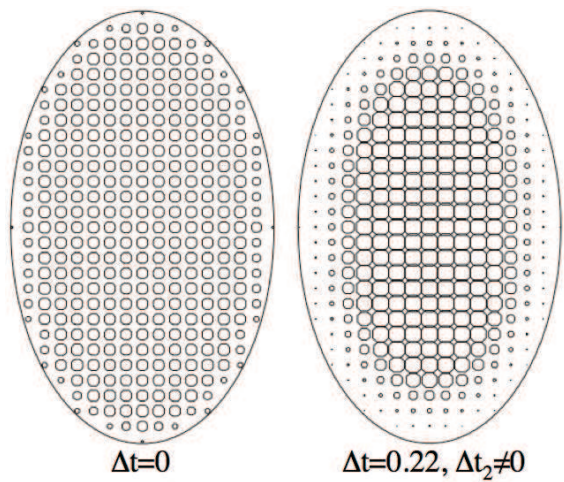}}
  \caption{Effect of shape on the charge expulsion profile. $t_h=0.1$, $U=2$.  In the ellipsoidal shape shown, for the conventional Hubbard model (left) 
  the charge occupation near the surface is
  similar near the top surface and the lateral surfaces. For the dynamic Hubbard model (right)  the hole concentration is somewhat higher near the lateral surfaces than near the top and
  bottom surfaces. For the origin of this effect, see text.
  Temperature is  $T=0.02$.}
\end{figure} 

\begin{figure}
\resizebox{8.5cm}{!}{\includegraphics[width=7cm]{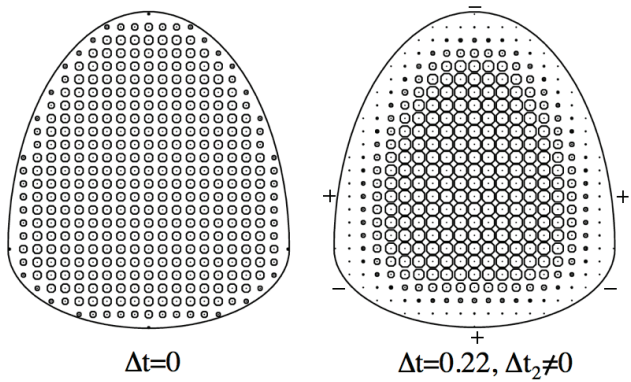}}
  \caption{Same as Fig. 10 for a case  where the upper half of the body is a prolate and the lower half an oblate ellipsoid of revolution.
  For the dynamic Hubbard model there is excess negative charge (lower hole concentration) 
  everywhere along the surface, however there is slightly more negative charge at the
  top than at the bottom, as indicated by the $-$ and $+$ signs. Variations along the lateral surfaces are also indicated with $-$ and $+$ signs.}
\end{figure} 

A qualitative way to understand this charge distribution in the superconducting state is the following: the electrons near the lateral surfaces can move faster than those
at the top and bottom in the region of high curvature, just as racing cars. Hence they will have higher kinetic energy and consequently lower potential energy than the
electrons near the top and bottom surfaces, so as to keep the same sum of kinetic and potential energies. Electrons near the lateral surfaces having lower potential energy
means that the electric potential is higher near the lateral surfaces than near the top and bottom surfaces, resulting in electric field lines starting from the side and
ending at the top and bottom just as found from the analysis of the hole distribution in the dynamic Hubbard model discussed in the above paragraph.

More generally, using these same arguments we expect that for other body shapes the electric potential near the surface will be higher in the regions of lower surface curvature and lower
in the regions of higher surface curvature in the dynamic Hubbard model and in superconducting bodies. An example of the charge distribution
 for a body shape resulting from combining
a prolate and an oblate ellipsoid is given in Figure 11. Examining the hole concentration in the various regions near the surface
for the right panel (dynamic Hubbard model)  it is seen for example that it is slightly higher
near the bottom surface that has a lower curvature, than near the top surface. The resulting charge profile varies as shown by the $+$ and $-$ symbols in the 
figure. This is qualitatively the same pattern that was found in  Ref. \cite{100} for the electric potential for 
a body of this shape by solving the modified electrodynamic equations   in the superconducting state.

\section{charge expulsion in the superconducting state}

As seen in the previous sections, the dynamic Hubbard model has a tendency to expel negative charge from its interior to the surface driven by lowering of
kinetic energy. Starting with a charge neutral system in the normal state, where a uniform positive
ionic charge distribution is compensated by an equal uniform  negative electronic charge distribution, the negative charge expulsion 
would result in a net charge distribution as qualitatively shown in Fig. 3: a net excess positive charge in the interior and
net excess negative charge near the surface. According to the numerical results in the previous sections (e.g. Fig. 4) the positive charge in the interior predicted by
the dynamic Hubbard model Hamiltonian is approximately uniform, just as that predicted from the electrodynamic equations in the superconductor\cite{chargeexp}.

This would result in the presence of an electric field in the interior of the system, that increases linearly in going from the center towards the surface.
However, this cannot happen in a real normal metal since a metal in the absence of current flow 
cannot have a macroscopic electric field in the interior. Therefore, we conclude that longer range
Coulomb interactions, omitted in the dynamic Hubbard model, prevent this from actually taking place in a real material. In other words, potential energy triumphs over
kinetic energy in the normal state, and a macroscopic metal will remain charge neutral in the interior, despite the $tendency$ to develop this macroscopic
charge inhomogeneity if dynamic Hubbard model physics is dominant. At most, the system will develop local charge inhomogeneity that will be screened
within a Thomas Fermi length, that can be several $\AA$ in systems like underdoped high $T_c$ cuprates where the carrier density is very low.

However, the situation can change if the system enters the superconducting state at low temperatures. There is no a-priori reason why a superconductor cannot
have an electric field in its interior\cite{chargeexp}. A superconductor is a macroscopic quantum system, and quantum systems in their ground state
minimize the sum of potential and kinetic energies. That should not in general result in a uniform charge distribution that minimizes
potential energy only. The electrodynamic equations that we have proposed for superconductors\cite{chargeexp} predict that the superconductor
has rigidity in the charge degrees of freedom\cite{chargeexp,rigidity} and will not screen an interior electric field as a normal metal would.

\begin{figure}
\resizebox{8.5cm}{!}{\includegraphics[width=7cm]{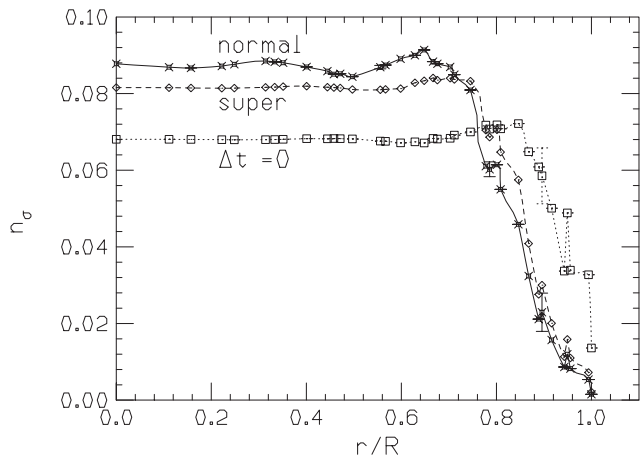}}
  \caption{ Comparison of occupations in the normal and superconducting states. 
  Radius of cylinder is $R=9$, average occupation per site is $n=0.126$. $t_h=0.1$, $U=2$, $\Delta t=0.2$, $\Delta t_2=0.4$, $k_B T=0.02$. 
  The average gap parameters in the superconducting state are $\Delta_{ii}=0.0044$, $\Delta_{ij}=-0.025$.
  We also show the occupations for the conventional Hubbard model, $\Delta t=0=\Delta t_2=0$. The charge expulsion is largest in the {\it normal state} of the dynamic
  Hubbard model.}
\end{figure}

\begin{figure}
\resizebox{8.5cm}{!}{\includegraphics[width=7cm]{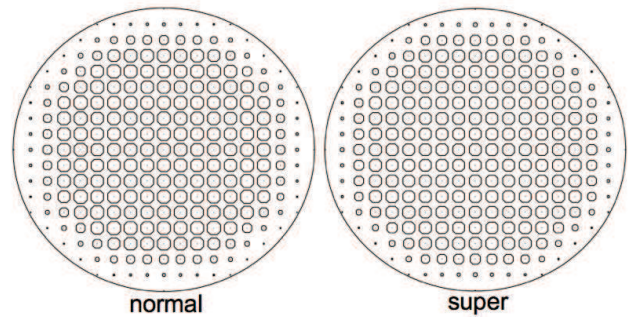}}
  \caption{Real space hole occupation for the cases of Fig. 12 with $\Delta t \neq 0$. Note that the interior hole occupation is slightly $larger$ in the normal state, and
  the occupation near the surface slightly smaller.}
\end{figure}

To compute the charge distribution in the superconducting state we solve numerically the Bogoliubov de Gennes (BdG) equations for the dynamic Hubbard model,
for systems with the same geometry as discussed in the previous sections. For the correlated hopping model
($\Delta t_2=0$) the equations are given in Ref. \cite{local}, and are simply extended for the case $\Delta t_2\neq 0$. There are two gap parameters,
$\Delta_{ii}$ and $\Delta_{ij}$ corresponding to on-site and nearest-neighbor pairing amplitudes\cite{local}.

\begin{figure}
\resizebox{8.5cm}{!}{\includegraphics[width=7cm]{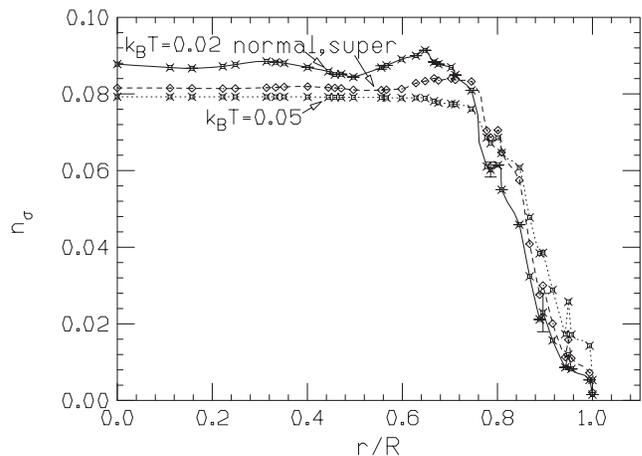}}
  \caption{Dynamic Hubbard model, parameters as in Fig. 12. Comparison of occupations right above $T_c$, $k_BT=0.05$, and at $k_BT=0.02$,
  with the system in the normal (full line) and superconducting (dashed line) state. Note that when the system is allowed to go superconducting,
  the occupations below $T_c$ essentially don't change, while if kept in the normal state the charge expulsion increases as the
  temperature is lowered.}
\end{figure} 

We have tested our computer program by solving the BdG equations numerically on a square lattice with periodic boundary conditions and comparing with the
standard BCS solution. For the cylindrical geometry with open boundary conditions considered here, the numerical solution obtained for the gap parameters deep in the 
interior are close to the gap parameters found in the square lattice with periodic boundary condition using both the BdG equations and the standard BCS equations
applicable to translationally invariant systems. We find that the gap parameters go to zero as the surface is approached. This agrees with what was   found 
by others\cite{tanaka} in a model
with no electron-hole asymmetry in the regime of low carrier density. Here we only study the low density regime and in addition this tendency is enhanced because
of the charge expulsion. 

Initially we had hoped\cite{chargeexp0} that comparison of the occupations in the dynamic Hubbard model in the normal and superconducting states would yield clear evidence that the
system expels negative charge from the interior to the surface as it goes superconducting,  as expected on physical grounds\cite{chargeexp0,giantatom} and
predicted by the electrodynamic equations\cite{chargeexp}. This is $not$ what happens, as shown in Figs. 12 and 13. Instead, the charge distribution becomes
{\it more uniform} in the superconducting compared to the normal state at the same temperature. In fact, it appears that as the temperature is lowered and
the system enters into the superconducting state the charge expulsion that increases in the dynamic Hubbard model as the temperature is lowered in the
normal state stops changing and stays essentially the same as what it is at $T_c$ when the system is cooled below $T_c$, as shown in Fig.14.

In summary, from the numerical results presented here it appears that 
the BCS/BdG solution of the dynamic Hubbard model does not reflect the charge expulsion predicted by the electrodynamic equations as the system enters
the superconducting state\cite{chargeexp}. On the other hand this is perhaps not too surprising. The charge expulsion predicted by the electrodynamic equations
is of the order of 1 extra electron every $10^6$ sites near the surface\cite{electrospin}, which certainly would not be noticeable in systems of the size considered here.
We have recently proposed that this predicted macroscopic charge inhomogeneity in the superconducting state should be
experimentally observable through the technique of electron holography\cite{holo1,holo2,holo3,holo4}.

\section{superconductivity and charge imbalance}

Within the BCS formalism, the total electronic charge per site is given by
\beq
Q_{tot}=\frac{2}{N} \sum_k [u_k^2 f(E_k)+v_k^2(1-f(E_k))]
\eeq
in units of the charge of one carrier, $e$ or $-e$ depending on whether one is using electron or hole representation. Eq. (13) can be written as\cite{qstar}
\beq
Q_{tot}=Q_c+Q^*
\eeq
with
\bmath
\beq
Q_c=\frac{2}{N}\sum_kv_k^2
\eeq
the charge of the condensate, and
\beq
Q^*=\frac{2}{N}\sum_k (u_k^2-v_k^2)f(E_k)
\eeq
\emath 
the average charge of the quasiparticles. The coherence factors are given by the usual form
\bmath
\beq
u_k^2=\frac{1}{2} (1+\frac{\epsilon_k-\mu}{E_k})
\eeq
\beq
v_k^2=\frac{1}{2}(1-\frac{\epsilon_k-\mu}{E_k})  .
\eeq
\emath
In a conventional BCS superconductor $Q^*=0$ in equilibrium since quasiparticles are charge neutral on average, half electron, half hole. A non-zero
$Q^*$, termed ``charge imbalance'' or ``branch imbalance'', can be generated in a non-equilibrium situation in the presence of current flow\cite{clarke,tc}
and/or a temperature gradient\cite{ps,ct}.

\begin{figure}
\resizebox{6.5cm}{!}{\includegraphics[width=7cm]{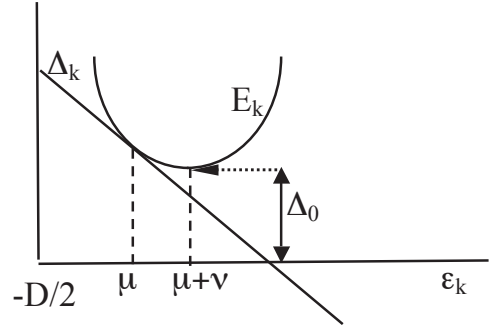}}
  \caption{Gap function $\Delta_k$ and quasiparticle energy $E_k$ as a function of band energy $\epsilon_k$  in  hole representation.}
\end{figure} 

Instead, in the dynamic Hubbard model (or the correlated hopping model) a branch imbalance exists even in equilibrium\cite{thermo}. 
The gap function has a slope of universal sign\cite{deltat}
 \beq
 \Delta_k\equiv \Delta(\epsilon_k)=\Delta_m(-\epsilon_k/(D/2)+c)
 \eeq
with $D$ the bandwidth and $\Delta_m>0$ and $c$ obtained from solution of the BCS equations\cite{deltat}. 
The minimum gap is $\Delta_0=\Delta(\mu)/a$, with $a=\sqrt{1+(\Delta_m/(D/2))^2}$ and the quasiparticle energy is given by
\beq
E_k=\sqrt{a^2(\epsilon_k-\mu-\nu)^2+\Delta_0^2} .
\eeq
The minimum gap $\Delta_0$ is attained not at $\epsilon_k=\mu$ but at $\epsilon_k=\mu+\nu$, with 
\beq
\nu=\Delta_m(T)\Delta_0(T)/(aD/2)>0 .
\eeq
Both $\Delta_0$ and $\Delta_m$ go to zero at $T_c$ as $\sqrt{T_c-T}$ so $\nu$ goes to zero linearly as $T$ approaches $T_c$ from below. The gap function and quasiparticle excitation spectrum are shown schematically in Figure 15 in  hole representation. 
In equilibrium, quasiparticles are symmetrically distributed around the minimum located at  $\epsilon_k^0=\mu+\nu$ and as
a consequence $Q^*>0$, quasiparticles are positively charged on average. If we ignore band edge effects we have simply
\beq
Q^*=\frac{2\nu}{N}\sum_k\frac{f(E_k)}{E_k}
\eeq
which is given approximately by (again ignoring band edge effects)
\beq
Q^*=\sqrt{8\pi}\frac{\nu(T)}{Da}\frac{e^{-\beta\Delta_0}}{(\beta\Delta_0)^{1/2}} ,
\eeq
$Q^*$ is suppressed at low temperatures due to the exponential factor, peaks somewhat below $T_c$ and goes to zero at $T_c$. Numerical examples are shown in
ref. \cite{thermo}. However when the finite bandwidth is taken into account $Q^*$ remains positive at $T_c$ and above.

 \begin{figure}
\resizebox{8.5cm}{!}{\includegraphics[width=7cm]{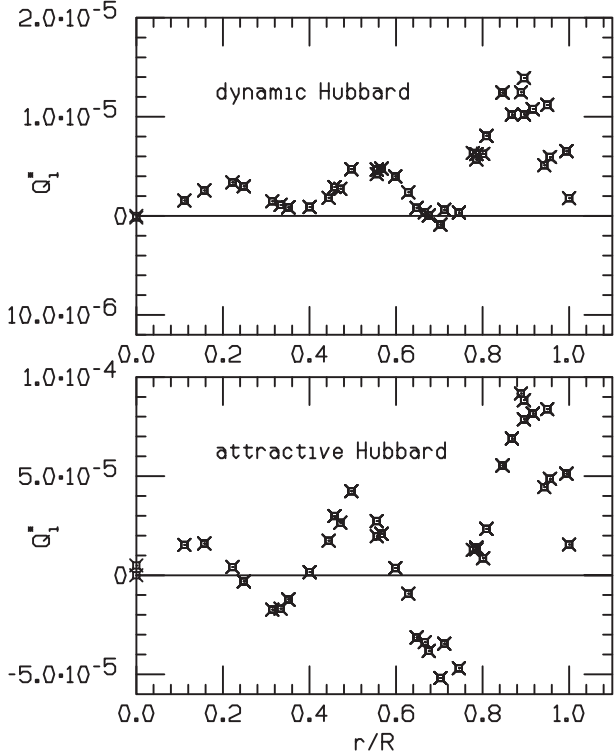}}
  \caption{ Quasiparticle charge (Eq. (35)) as function of distance to the center for a dynamic Hubbard model with parameters as in Fig. 12 and for an attractive Hubbard
  model with $t_h=0.1$, $U=-0.4$. Average occupation is $n=0.126$, temperature is $k_BT=0.01$. Note that in the dynamic Hubbard model the
  quasiparticle charge is predominantly positive.}
\end{figure}

Figure 16 shows the distribution as a function of the distance to the center $r$
of the quasiparticle charge $Q_i^*$ at site $i$, given by
\beq
Q_i^*=\frac{2}{N}\sum_n(u_{ni}^2-v_{ni}^2)f(E_n)
\eeq
 obtained from solution of the BdG equations, for a dynamic Hubbard model and for an attractive
Hubbard model. In Eq. (22), $u_{ni}$ and $v_{ni}$ are the amplitudes of the n-th eigenvector at site $i$  obtained from diagonalization of the BdG Hamiltonian\cite{local},
$E_n$ is the energy for state $n$ and $f$ is the Fermi function. 
In the attractive Hubbard model (Fig. 16 lower panel)  particle-hole symmetry is broken only because the band is not half-full, but the interaction is particle-hole symmetric.
As a consequence, the quasiparticle charge oscillates between positive and negative values. Instead, as seen in Fig. 16 (upper panel) in the dynamic Hubbard model quasiparticles are predominantly
positively charged, as expected due to the shift in the chemical potential by $\nu$ displayed in Fig. 15.

Figure 17 shows the real space distribution of the quasiparticle charge in the dynamic Hubbard model (right panel). The total site occupation for this case
is shown on the left panel. 
It can be seen from Figs. 16 and 17 that the positive quasiparticle charge is located mostly near the surface of the system. This is relevant to the discussion in the next section.

 \begin{figure}
\resizebox{8.5cm}{!}{\includegraphics[width=7cm]{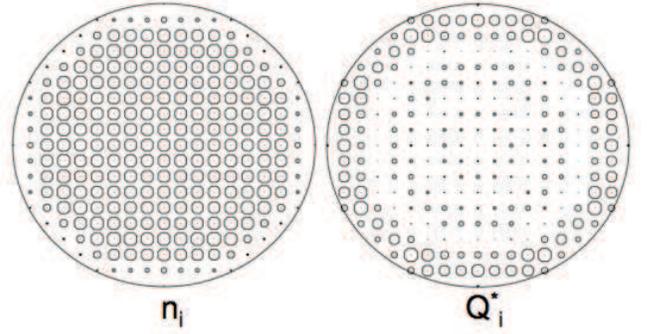}}
  \caption{Real space hole occupation (left panel) and quasiparticle charge (right panel) in the superconducting state of the dynamic
  Hubbard model. Parameters are the same as for Figure 16 (upper panel). Note that the quasiparticle charge is predominantly near the
  surface of the sample.}
\end{figure}

\section {two-fluid model and interior electric field}
 In this section we discuss to what extent the dynamic Hubbard model reflects the physics shown in Figure 3 in the superconducting state, how it depends on
temperature, and how this physics could be
detected experimentally.

We assume a two-fluid model, with the total carrier concentration independent of temperature. We have then
\beq
n_s=n_s(T)+n_n(T)
\eeq
with $n_s$ the total carrier (hole) concentration and $n_s(T)$ and $n_n(T)$ the superfluid and normal components at $0\leq T\leq T_c$.
Within the two-fluid interpretation of BCS theory they are given in terms of the London penetration depth as
\bmath
\beq
n_s(T)=n_s\frac{\lambda_L^2}{\lambda_L^2(T)}
\eeq
\beq
n_n(T)=n_s\lambda_L^2(\frac{1}{\lambda_L^2}-\frac{1}{\lambda_L(T)^2})
\eeq
\emath
with $\lambda_L$ ($\lambda_L(T)$) the London penetration depths at zero (finite) temperature. Ignoring finite bandwidth effects the quasiparticle
density is then\cite{tinkham}
\beq
n_n(T)=2n_s\int_{\Delta_0}^\infty dE (-\frac{\partial f}{\partial E})\frac{E}{\sqrt{E^2-\Delta_0^2}}
\eeq
which yields at low temperatures
\beq
n_n(T)=\sqrt{2\pi}(\beta\Delta_0)^{1/2}e^{-\beta \Delta_0} n_s .
\eeq
On the other hand, the average quasiparticle charge per site is given by Eq. (21). Combining with Eq. (26),
\beq
\frac{Q^*(T)}{n_n(T)}=\frac{1}{a}\frac{k_BT}{D/2}\frac{\nu}{\Delta_0}=\frac{k_B T \Delta_m}{\Delta_m^2+(D/2)^2}\sim\frac{k_BT\Delta_m}{(D/2)^2} .
\eeq
It can be seen that the quasiparticle charge is a small fraction of the quasiparticle density. For example, for the parameters of Fig. 16 we have
\beq
D=2z(t_h+n\Delta t+\frac{n^2}{4}\Delta t_2)=1.01 ,
\eeq
$\Delta_{ij}=0.00253$, $\Delta_m=z\Delta_{ij}=0.10$, hence $Q^*/n_n=0.004$.

Assuming the system as a whole is charge neutral, the negative charge of the electrons in the band exactly compensates the positive charge of the
ions, which is uniformly distributed in space (except for variations on the scale of $\AA$). At temperatures below $T_c$, the quasiparticles have a
net positive charge, hence as a consequence the condensate has a total negative charge $greater$ than the total positive charge of the ions. The condensate is
highly mobile, and just as in a normal metal any excess negative charge will move to the surface\cite{thomson} it is natural to expect that 
negative charge from the condensate will move to the surface.

Furthermore, we have seen in the previous section that the positive quasiparticle charge is located predominantly near the surface in the
superconducting state (Fig. 17 right panel). This can be interpreted as reflecting the fact that the superfluid has higher negative density near the surface,
and the positive  normal fluid consequently develops higher density near the surface to screen the superfluid charge. In addition, as already seen in the normal state of the dynamic Hubbard
model, the total hole concentration is smaller near the surface which implies extra negative charge near the surface. Thus we argue that the
dynamic Hubbard model provides support to the prediction of the 
electrodynamic equations of the theory\cite{chargeexp0,chargeexp} that the superconductor expels superfluid negative charge
from the interior to the surface.

Whether or not a macroscopic electric field will exist in the interior of the superconductor depends on whether there are enough quasiparticles to screen the
electric field created by the negative charge expulsion of the condensate. In the ground state (no quasiparticles) the theory predicts that the
net positive charge density in the interior is\cite{electrospin}
\beq
\rho_0=\frac{r_q}{R} |e| n_s=\rho_s+\rho_{ions}
\eeq
with $R$ the radius of the cylinder and $r_q=\hbar/(2m_ec)=0.00193 \AA$ the quantum electron radius, and there is a negative charge density
\beq
\rho_-=-\frac{R}{2\lambda_L}\rho_0=\rho_s+\rho_{ions}
\eeq
within a London penetration depth of the surface, as shown schematically  in Fig. 3. The charge densities at zero temperature are  shown schematically in the left panels of Figure 18. $\rho_s$ denotes the superfluid charge density.

 \begin{figure}
\resizebox{8.5cm}{!}{\includegraphics[width=7cm]{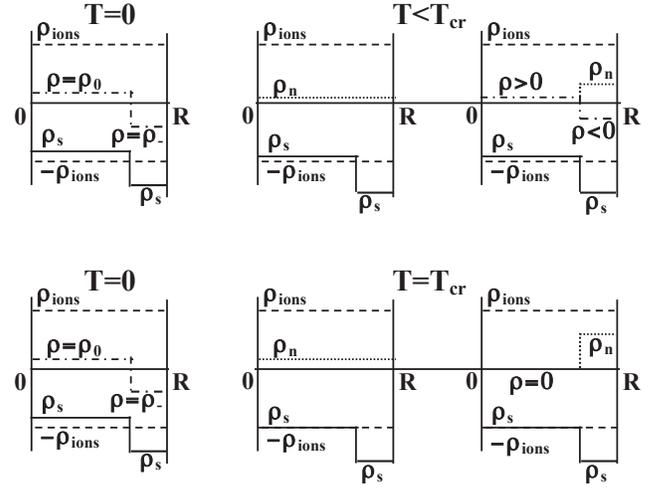}}
  \caption{Charge distribution in the superconductor (schematic). In the ground state (left panels) there is an excess positive charge density $\rho_0$ in the interior and an
  excess negative charge density $\rho_-$ near the surface. At low temperatures $T<T_{cr}$ (upper panels) the charge density is partially screened by excited
  quasiparticles but a net positive charge density $\rho>0$ remains in the interior and a net negative charge density $\rho<0$ near the surface 
  (dot-dashed lines in upper right panel). At temperature $T=T_{cr}$ and above
  the net charge density  is zero both in the interior and near the surface.}
\end{figure} 

At finite temperatures, there is a positive quasiparticle charge density excited, $\rho_n=|e|Q^*$, and a crossover temperature $T_{cr}$ can be defined. 
For temperatures lower than $T_{cr}$, the average quasiparticle charge density excited is smaller than $\rho_0$. In the middle top panel of Fig. 18 we assume
$\rho_n$ is uniformly distributed, and in the right top panel we assume all $\rho_n$ has moved to within the London penetration depth of the surface. Even so,
it is unable to screen the internal electric field, since a positive net charge density $\rho=\rho_0-\rho_n>0$ remains in the interior and a negative net charge
density $\rho<0$ remains near the surface, as shown in the top right panel of Fig. 18. At the crossover temperature $T_{cr}$ the quasiparticle charge density excited  
reaches the value $\rho_0$, and by migrating to the region within a London penetration depth of the surface (lower right panel of Fig. 18) it can completely
screen both the interior positive charge and the negative charge in the surface layer, so that the electric field everywhere gets cancelled.
This will also be the case at any temperature $T>T_{cr}$.

The value of the crossover temperature can be obtained from the equation
\beq
Q^*=\frac{Q^*}{n_n}n_n=\frac{r_q}{R}n_s
\eeq
with $Q^*/n_n$ given by Eq. (27) and $n_n$ given by Eq. (26), hence
\beq
\frac{k_B T_{cr}\Delta_m}{(D/2)^2}(\frac{2\pi\Delta_0}{k_BT_{cr}})^{1/2}e^{-\Delta_0/(k_BT_{cr})}=\frac{r_q}{R} .
\eeq
For example, assuming the usual relation $2\Delta_0/k_BT_c=3.53$, 
for the case under consideration with $Q^*/n_n=0.004$ and assuming
$R=500 nm$ yields $T_{cr}=0.16T_c$. For temperatures lower than $T_{cr}$, a nonzero electric field is predicted to exist in the interior of the superconductor.

In the presence of a non-zero internal electric field, superconductors of non-spherical shape should also develop electric fields extending 
to the region exterior to the body, of magnitude and direction determined by the shape of the body and the electrodynamic equations of the 
superconductor\cite{ellipsoid,100}. These electric fields should be  experimentally detectable in the neighborhood of superconductors at
temperature lower than $T_{cr}$.
In addition, the internal electric field should be directly detectable in electron holography experiments\cite{holo1,holo2,holo3}.
The magnitude of these predicted electric fields is of order of $H_{c1}$, the lower critical magnetic field, in the interior of the superconductor\cite{electrospin}, and an appreciable
fraction of it in the region outside the superconductor near the surface, depending on the shape of the body\cite{ellipsoid,100}. No external electric field is
expected outside a planar surface or a spherical body.

\section{  the Meissner effect, the London moment and the gyromagnetic effect}

The fact that in superconductors the superfluid carries {\it negative charge} is established experimentally from experiments that measure the London moment\cite{lm1,lm2}:
 a rotating superconductor develops a magnetic field that is always {\it parallel}, never antiparallel, to the direction of the mechanical angular
momentum of the body\cite{ehasym}.

We have seen that the dynamic Hubbard model has a tendency to expel negative charge, that in a real system is inhibited in the normal state because of the effect of 
long-range Coulomb repulsion but may take place when the system becomes superconducting. The considerations in the previous section suggest that as
a system becomes superconducting the electrons that condense into the superfluid state are partially expelled towards the surface, and at the same time
normal electrons flow inward to compensate for the charge imbalance, as indicated by the fact that the positive quasiparticle charge moves outward in the
superconducting state as seen in the last section.

Consider now these processes in the presence of an external magnetic field in the $z$ direction, as shown in Fig. 19. The outflowing superfluid electrons will be deflected
counterclockwise by the Lorentz force exerted by the magnetic field, building up a Meissner current flowing clockwise near the surface that suppresses the
applied field in the interior.
At the same time, the inflowing normal electrons are deflected clockwise by the magnetic field. Because these electrons undergo scattering from the ions, they will transmit their
momentum to the ions and the body as a whole will start rotating in a clockwise direction. And because these electrons are slowed down and ultimately stopped
by the collisions
with the ions they will not  reinforce the applied magnetic field. The end result is a superfluid current near the surface flowing in clockwise direction
(i.e. superfluid electrons flowing in counterclockwise direction) that
suppresses the interior magnetic field, and a slow body rotation in the clockwise direction that exactly cancels the mechanical angular momentum carried by the
superfluid electrons in the Meissner current, as required by angular momentum conservation.

 \begin{figure}
\resizebox{8.5cm}{!}{\includegraphics[width=7cm]{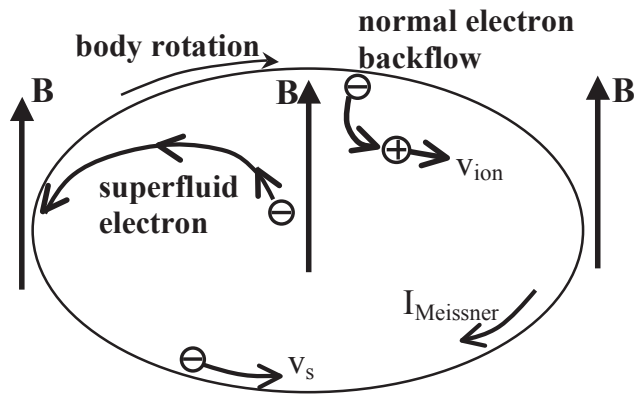}}
  \caption{The outflowing superfluid electrons are deflected in the counterclockwise direction by the applied magnetic field, generating 
  a clockwise Meissner current ($I_{Meissner}$)
  near the surface that suppresses the magnetic field in the interior. The inflowing normal electrons collide with the ions and impart the entire body with
  angular momentum antiparallel to the applied magnetic field, that is equal and opposite to the mechanical angular momentum of the
  electrons in the Meissner current.}
\end{figure}

The magnetic field generated by rotating superconductors (London moment)\cite{lm1,lm2} can be similarly explained\cite{spincurrent} by the fact that in a rotating normal metal that is cooled
into the superconducting state the expelled superfluid electrons, that are moving at the same angular velocity as the body,  will have a smaller tangential
velocity than the body when they reach the surface, giving rise
to a net current and resulting magnetic moment in direction $parallel$, never antiparallel, to the angular momentum of the rotating body.

More quantitatively, the outflow occurs because superfluid electrons expand their orbits from microscopic radius $k_F^{-1}$ to mesoscopic radius 
$2 \lambda_L$\cite{sm}, in the process acquiring an azimuthal velocity\cite{azim}
\beq
v_\phi=-\frac{e}{2m_ec}r=-\frac{e}{m_ec}\lambda_L B
\eeq
which is the speed of the superfluid electrons in the Meissner current\cite{tinkham}. The total mechanical angular momentum acquired by these electrons in a cylinder
of radius $R$ and height $h$ is
\beq
L_{el}=\pi R^2 h n_s (m_e v_\phi 2 \lambda_L)
\eeq
which coincides with the total electronic angular momentum of the Meissner current flowing within a London penetration depth of the surface
\beq
L_{Meissner}=2\pi R \lambda_L h n_s (m_e v_\phi R).
\eeq
The inflowing normal electrons transmit the same angular momentum to the body as a whole by collisions with the ions, as required by angular momentum
conservation\cite{angmom1}. As a consequence, the body starts rotating with angular velocity determined by the condition of angular momentum conservation,
as measured experimentally\cite{gyro1,gyro2,gyro3} (gyromagnetic effect).
 
These processes provide an intuitive explanation for the dynamics of the Meissner effect\cite{sm}, the generation of the London moment, the 
gyromagnetic effect and the
puzzle of angular momentum conservation\cite{angmom1,angmom2} in superconductors, $provided$ the superconductor is  described by a dynamic Hubbard model
that gives rise to negative charge expulsion.
 In contrast, in superconductors  not described by dynamic Hubbard models
but by conventional BCS-electron-phonon  theory\cite{tinkham} no charge expulsion takes place  and hence these considerations don't apply. For those superconductors,
if they exist, the 
dynamical origins of the Meissner effect and the London moment and the explanation of the angular momentum puzzle remain to be elucidated.

\section{discussion}
Both the conventional Hubbard model and the dynamic Hubbard model are simplified descriptions of real materials,  and whether or not they contain
the physics of interest for particular real materials is in principle an open question. In this paper we have argued that the new physics that
the dynamic Hubbard model incorporates beyond what is contained in the conventional Hubbard model is key to understanding many properties of high $T_c$ cuprates as well as
of superconductors in general.

The new physics of the dynamic Hubbard model   is that it allows the electronic orbital to expand when it is doubly occupied, as it occurs in real atoms. This expansion has
associated with it outward motion of negative charge as well as lowering of the electron's kinetic energy {\it at the atomic level}, and it is of course
electron-hole $asymmetric$ (the orbital does not change when a second $hole$ is added to
a non-degenerate orbital occupied by one hole). In the conventional Hubbard model instead,
the state of the first electron in the orbital does not change when a second electron is added,
the kinetic energy of the electron does not change (the potential energy does), and the model
is electron-hole symmetric.

We have argued in this paper and in previous work that these three properties that 
occur already at the atomic level in the
 dynamic Hubbard model, negative charge expulsion, lowering of 
electronic kinetic energy, and electron-hole asymmetry, 
 are key to understanding high $T_c$ superconductivity in the
cuprates and superconductivity in general. Furthermore we have shown
in this paper that these properties are also  displayed by the entire system described by
a dynamic Hubbard model {\it in the normal state}.

The tendency of the dynamic Hubbard model to expel negative charge and the tendency to pairing of holes and superconductivity driven by kinetic energy lowering
of course go hand in hand: 
they both originate in the fact that the kinetic energy of a hole is lowered when another hole is nearby. Increasing the chance of having another hole nearby can
be achieved by negative charge expulsion, thus increasing the $overall$ hole density, and by pairing, thus increasing the $local$ hole density.
The propensity of dynamic Hubbard models to develop charge inhomogeneity and 
the high sensitivity to disorder of the local superconducting gap found in earlier work\cite{local}
also go hand in hand, since both originate in the fact that the kinetic energy varies with
charge occupation, which is what gives rise to a kinetic-energy-dependent
pair interaction and superconducting gap function\cite{deltat}.

We restricted ourselves in this paper to the antiadiabatic limit, i.e. assuming that
the energy scale associated with the orbit expansion ($\omega_0$ in Eq. (4)) is sufficiently large than it can be assumed infinite. This brings about
the simplification that the high energy degrees of freedom can be eliminated and the Hamiltonian becomes equivalent to the low energy
effective Hamiltonian Eq. (7), a Hubbard model with correlated hoppings, linear and nonlinear terms $\Delta t$ and $\Delta t_2$. 
This low energy effective Hamiltonian, together with the quasiparticle weight renormalization described by Eq. (6), describes many properties
that we believe are relevant to real materials and are not described by the conventional Hubbard model. In particular  it gives rise to hole superconductivity\cite{deltat},
driven by lowering of kinetic energy of the carriers\cite{apparent,kinetic},
with many characteristic features that resemble properties of the cuprates. 
In other work we have also examined the effect of the high energy degrees of freedom in describing spectral weight transfer from 
high to low energies (``undressing''\cite{undr}) as the number of holes increases and as the system enters the superconducting state,
as well as the effect of finite $\omega_0$ in further promoting pairing in this model\cite{dynfrank}.

The effects predicted by this Hamiltonian are largest when the coupling constant $g$ is large, or equivalently when  the overlap matrix element
$S$ is small, which corresponds to a ``soft orbital'' that would 
exist for negatively charged anions, $and$ the effects are also  largest when the band is almost full with negative electrons
(strong coupling regime)\cite{strong}. Thus, not surprisingly, more negative charge at the ion  
or/and in the band yield  larger tendency to negative charge expulsion for  the
entire system.
 We believe that the Hamiltonian is relevant to describe the physics of   superconductors  
including high $T_c$ cuprates, $Fe$ pnictides, $Fe$ chalcogenides, $MgB_2$ and $BiS_2$-based\cite{bis2} superconductors. These materials have
negatively charged ions ($O^{--}, As^{---}, S^{--}, Se^{--}, B^-$) with soft orbitals, and for most of them, including ``electron-doped'' cuprates\cite{edoped}, there is 
experimental evidence for dominant $hole$ transport in the normal state. We
  suggest that the orbital expansion and contraction of these
negative ions depending on their electronic occupation is responsible for many interesting properties of these materials including their superconductivity,
and is described by the dynamic Hubbard model.

We have also examined here  the question whether the interior\cite{chargeexp} and exterior\cite{ellipsoid} electric fields predicted to exist in the ground state of superconductors 
within this theory  would exist also at finite temperatures and  concluded that they should exist and hence be
experimentally detectable up to a crossover
temperature $T_{cr}$, calculated to be $0.16T_c$ in one example.
We also examined   the effect of the body shape (surface curvature) on the charge distribution near the surface in the normal state of the model and found that
it is consistent with the pattern of electric field dependence on particle shape predicted in the superconducting state\cite{ellipsoid,100}.

Finally, we have proposed that the negative charge expulsion predicted by dynamic Hubbard models is relevant to the understanding of the 
Meissner effect, the London moment and the gyromagnetic effect exhibited by all  superconductors.

In summary, we argue that it is remarkable that the dynamic Hubbard model exhibits the same physics at the level of the single atom and of the system as a whole,
and both in the normal and in the superconducting states, and that the same physics is found in different ways through seemingly different 
physical arguments and mathematical equations. In particular, we hope the reader will appreciate the remarkable qualitative similarity of Figs. 1, 2  and 3,  depicting 
the charge distribution in an atom, 
a system in the normal state and
a superconductor within this model. Superconductors have been called ``giant atoms'' in the early days of 
superconductivity for many $other$ reasons\cite{ga1,ga2,ga3}. The essential property of the atom, that it is $not$ electron-hole symmetric because the 
negative electron is much lighter than the positive nucleus, manifests itself   in the atom described by the dynamic Hubbard model and in the
state of a macroscopic superconducting body  described by the model, and is missed in the world described by particle-hole symmetric conventional
Hubbard or Fr\"{o}hlich models both at the atomic level and at the level of the macroscopic superconductor.
The superconductor closely resembles a ``giant atom'' within our description, with the highly mobile light negative superfluid reflecting the   atomic electron, and the
heavy positive quasiparticles reflecting the positive nucleus.

Much of the physics of   dynamic Hubbard models for finite $\omega_0$ remains to be
understood. In fact, the model itself may require substantial modification to account for different values of $\omega_0$ for different electronic occupations:
the excitation spectrum of the neutral hydrogen atom, $H$,  is certainly very different from that of $H^{-}$. In connection with this and 
going beyond the antiadiabatic limit where only diagonal transitions of the auxiliary boson field are taken into account as in this paper,
it is possible that $vertical$ transitions may play a key role in describing the superconducting state\cite{eh3}. It is also an open question to what extent dynamic Hubbard models can
describe the mysterious ``pseudogap state'' of underdoped high $T_c$ cuprate materials. These and other questions will be the subject of future work.

 \end{document}